\documentstyle[preprint,aps,twocolumn,epsf]{revtex}
\begin{document}
\draft
\title{ Raman scattering in high temperature superconductors :
An Integrated view}
\author{S. N. Behera$^\star$, Umesh A. Salian and Haranath Ghosh.}
\address{Institute of Physics,
Bhubaneswar 751 005, Orissa, India.}
\maketitle
\abstract{ 
The common features in the Raman data of high temperature
superconductors: (the cuprates, bismathates, alkali doped
fullerides and some organic superconductors), are analyzed. It was
shown that qualitative understanding of the data can be achieved
in terms of non-Fermi liquid models for their normal state, with 
appropiate bag mechanisms for the superconducting state.
}
\narrowtext


\section{Introduction}
Over the years Raman scattering has proved to be a valuable tool
for probing the normal as well as the superconducting (SC)
states of the high temperature superconductors. It is well known
that in the cuprate superconductors the measured Raman
intensities have shown several unusual features \cite{1}, 
which are summarized
elsewhere \cite{2,3,4}. Similar behaviour of the Raman spectra
has also been observed in other families of high temperature
superconductors such as the bismuthates \cite{5} the alkali
doped fullerides \cite{6} and the organic superconductors
\cite{7}. For example the Raman spectra of BKBO ($Ba_{0.6}K_{0.4}BiO_3$) in
the metallic state shows two phonons out of which the lower
frequency one developes an assymetric line shape and the other
with higher frequency softens on lowering the temperature.
Furthermore, a new peak appears below certain temperature which
persists in the SC state. In contrast these features are not
present in the insulating samples. On the other hand in case of
alkali doped fullerides and the organic superconductor
$(BEDT-TTF)_2I_3$ the intensities of some of the phonons are
suppressed in the metallic state or on going to the SC state respectively.
Again in all these systems except for the alkali doped
fullerides a constant intensity background has been
reported, which is attributed to scattering by charge carriers.
\par The observation of these global features in the Raman
spectra of the different families of the high temperature
superconductors is rather surprising. The fact that
these systems exhibit entirely different structures and
properties suggests that the normal state and their
superconductivity may have entirely different origin. For
example the cuprates and the organic SC are two dimensional
systems whereas the bismuthates and the alkali doped fullerides
are three dimensional in nature. Similarly the cuprates 
exhibit antiferromagnetic order 
and are Mott insulators in their undoped state whereas the
bismuthates are well known to be charge density wave (CDW)
insulators \cite{8}. There is some evidence (e.g. existence of 
nested pieces of 
Fermi surface) in favour of the CDW state being the
ground state in the case of the other two systems as well
\cite{9,10}, namely the alkali doped fullerides and the organic
superconductors.
\par The other unexpected feature is the observation of the
constant intensity background in the Raman spectra in almost all
the high $T_c$ systems, which was never observed before for a
usual metal. 
The charge carrier scattering essentially
involves the polarizability which for a Fermi liquid like metal
is given by the Lindhard function. Since the inelastic light
scattering takes place with no momentum transfer, the Lindhard
function vanishes at q = 0, hence there is no scattering by
charge carriers in a metal. 
Again  the phonon self
energy in a usual metal 
will also vanish for the zone centre phonon , prohibiting
any observable change in the phonon peak. 
The very fact that in the high $T_c$ systems 
charge carrier scattering as well as phonon anomalies has been observed
points to the fact that the normal
states of all these systems exhibit non-Fermi liquid like
behaviour. Furthermore, the unexpected frequency and temperature
dependence of the imaginary part of the polarizability, deduced
from the fact that the background intensity is constant
even indicates that the normal state of all these systems is
a marginal Fermi liquid \cite{11}. 
\par Following this dictum we have proposed (i) a strongly
correlated metallic state for the cuprates \cite{2,3} and (ii) a
fluctuating CDW state to be the ground state for the
bismuthates, the alkali doped fullerides and the organic
superconductor $(BEDT-TTF)_2I_3$. Both of these states have non-Fermi
liquid excitation spectra, and a lot of other similar characteristics.
Superconductivity in both these systems arises because of a bag
mechanism \cite{2,3,8,9,10}. In what follows we shall provide
qualitative accounts of these models (Sec. 2) and calculate the
Raman scattering intensity (Sec. 3). Some results of recent
calculations are presented in Sec. 4 for the cuprates,
bismuthates and alkali doped fullerides, which are obtained by
adopting these models to the respective systems.
\section {Non-Fermi Liquid Normal States and Superconductivity}

The phase diagrams of the cuprates are dominated by an
antiferromagnetic ground state for the undoped system which is a
consequence of strong electron correlation. On doping with
charge carriers the system becomes a Mott insulator with short
range magnetic correlations. This state is proposed to be a
resonating valence bond (RVB) insulator \cite{12}, which on
further doping undergoes an insulator-metal transition. Doping
charge carriers, while turning the RVB insulator into a metal
also creates strong fluctuations in the RVB state, which
corresponds to the spontaneous breaking and making of these
bonds. The quanta of these fluctuations which are the collective
modes of the RVB state, therefore interact strongly with its elementary
excitations. This interaction in fact is the
most crucial ingradient of the strongly correlated metallic
state. The details of the correlated metallic state and its
consequence on Raman scattering are presented in \cite{2,3}. It
was shown that this correated metallic state exhibits marginal
Fermi liquid behaviour. The effect of the strong interaction
between the elementary excitations and the quanta of the
collective modes of the RVB state is taken into account
phenomenologically by assuming that as a consequence of this
interaction the RVB gap becomes a pseudo gap existing only in
some regions of the Fermi surface, leaving the Fermi surface
intact in other regions, as was originally suggested by Bilbro
and McMillan \cite{13} for low dimensional systems. With
increasing dopant concentration, the region in the Fermi surface
with the RVB gap shrinks. Ultimately when the gapped
region of the Fermi surface vanishes, which can happen when the
dopant concentration exceeds a critical value, there are no more
bonds left in the metallic state and the normal state admits the
usual Fermi liquid description and becomes a good metal.
However it is only in the correlated (bad) metallic state that
superconductivity appears \cite{4}. This provides a good description 
of a large number
of observed properties of the cuprates in the normal state. 
\par The bismuthate $(BaBiO_3)$ in its undoped form is known to
be a CDW insulator, hence a description almost similar to the cuprates
holds for the doped systems. On doping the system with 
impurities like K, holes are created in the fully occupied lower
CDW band. On increasing the hole concentration the CDW gap will
turn into a pseudo gap due to partial removal of nested pieces
of Fermi surface. Furthermore creation of holes will cause
strong fluctuation in the CDW order parameter, the quanta of
these being the collective modes. These collective modes in turn
will interact strongly with the elementary excitations of the
CDW state. This interaction will dominate the normal state of
the system, and when the dopant concentration exceeds a critical
value the nesting of the Fermi surface is totally destroyed so
the system falls back to the usual Fermi liquid like metal.
Again superconductivity appears only in the nested metallic
state. Thus the normal states of both the cuprates and the
bismuthates are very similar in nature and their elementary
excitations exhibit marginal Fermi liquid behaviour, even though
it is entirely of different origin in the two systems. This
explains why the Raman spectra in the two systems exhibit
similar features.
\par Superconductivity in these systems arises because of the pairing
of the corresponding elementary excitations due to a bag
mechanism \cite{14,8}. In either case the elementary excitations
are strongly interacting with the collective
modes of the system. Therefore, there can be a collective mode
mediated attraction between the elementary excitations which
will give rise to pairing. In the original bag models
\cite{14,8} only the pairing of the holes in the lower band was
considered. However in the present case since there exists only a
pseudogap in these system there can be pairing of elementary
excitations belonging to both the valence and conduction bands
(lower and upper bands). This is the modified charge/correlation 
bag model for SC in the bismuthates/cuprates, which
explains the experimental Raman data much better \cite{3,10} as
will be discussed below. 
\par The fluctuating CDW metallic state and the modified charge
bag model can also be adopted to the alkali doped fullerides
\cite{9} and the organic superconductors \cite{10}.
\section {Results For Raman Intensities}
The total Raman
intensity in general is proportional to \cite{4},
\begin{eqnarray}
I(q,\omega)&=&[1+n(\omega)](\sum_i [-Im\chi_i(q,\omega)] \nonumber \\
& &
+\sum_j
[Im D_j(q,\omega)])
\end{eqnarray}
where $\chi_i(q,\omega)$ are the wave vector (q) and frequency
$(\omega)$ dependent polarizabilities, $D_j(q,\omega)$ are the Green's
functions of the various Raman active phonons and $n(\omega)$ is the Bose
factor. It should be
noted that the same  polarizability functions which determine the
charge carrier scattering also provide the self energy to the
phonons. The polarizability functions have to be calculated for the normal
and the SC states in order to determine the Raman intensities.
As described earlier \cite{3,4,10,15} in these systems
there exist two different kinds of phonons which contribute to
Raman scattering, (i) the ususal zone centre (q=0) optic phonons
(the propagating ones) and (ii) the zone folded (q=Q) phonons
which arise due to the superperiodicity of the 
antiferromagnetic (CDW) order in the cuprates  (bismuthates). Since
in the doped systems the long range order is reduced to a local order, the
later category of phonons will be more localized in nature.
These two intereact differently with the charge carriers, while
the propagating ones couple to the charge carrier density (the
usual electron-phonon interaction) the localized ones interact
with the collective (amplitude) modes of the RVB/CDW condensate
 in the normal state. In calculating the polarizability
the density of states at the Fermi level is devided into two
parts N(0) = $N_1 (0)$ + $N_2$ (0), where $N_1$(0) corresponds
to the region of the Fermi surface with a gap in the normal state 
and $N_2$(0) to the region where the gap
vanishes. This is how the Bilbro-McMallian approximation is
implemented. With this prescription the Raman intensities are
calculated. 
\par In the case of the cuprates the results of a zero
temperature calculation for q=0, was presented in \cite{2},
while that of a finite temperature calculation for the
correlated metallic state as well as the correlation bag model
of superconductivity was reported in \cite{3}.
\noindent The finite
temperature study involved a low and high temperature expansion of the
Fermi functions \cite{3}. Furthermore a small q, but
zero temperature calculation was carried out to account for the
penetration of the electromagnetic raditaion into the sample (the skin depth).
The cuprates being bad metals (extreme type II materials) in their
normal (SC) states the skin depths are rather large which
justifies the small $q$ expansion.
The calculated Raman intensities within this approximation was presented in
\cite{4} which shows qualitative agreement with the Raman data. The
finite q calculation also yields the phonon dispersion which is
plotted in Fig.1(a) for the propagating mode. The figure shows
the dispersion in the normal and the SC states \cite{16}. The
merger of two phonon branches in the normal state and the
splitting of a branch into two branches in the 
SC state are  noteworthy.
Correspondingly the peaks in the spectral density function
approach each other and merge(split) into one(two) peak beyond (below) a
certain wave vector. 
\par In the SC state the lowest frequency branch has very little intensity
as seen from the plot of the spectral density functions. Therefore the
difference in the phonon frequencies of the lower branch of the normal
state and the branch appearing just below it in the SC state was taken to
be the wave vector dependent shift. 
Figure 1(b) shows this wave vector dependent
softening of the phonon frequency of the lower branch on going
over to the superconducting state. Note that the shift in phonon
frequency with wave vector increases at first and then
saturates. This is in qualitative agreement with the results of
neutron scattering measurements \cite{17}. The Raman
intensity calculations for the cuprates was discussed at length
in \cite{2,3,4,15}. 
\par In the bismuthates the fluctuating CDW state 
and the modified charge bag model \cite{18} provide a good description of the
normal and the 
SC states respectively. \\

\vskip 0.3cm
\noindent {\small Fig 2(a). Calculated Raman intensity at three different 
temperatures using the temperature expansion approximation.} \\


 In the K-doped system two 
phonons have been  observed \cite{5}. The high frequency 
one is taken to be the propagating and the low frequency one to be the localized
phonon. The CDW gap lies between the two  phonons. The Raman
intensities are calculated at finite temperatures by two
different methods (i) by making high and low temperature
expansions, and (ii) by performing an all temperature calculation, both of
which are carried out numerically. 
The temperature expansion results for the normal state are 
shown in Fig(2a) for three different temperatures.
 It can be seen that in this approximation the low
frequency phonon looses its intensity drastically, while it
hardens and broadens as the temperature is lowered (increasing
$c1$). The high frequency phonon as well as the constant background
intensity are not effected very much with decreasing temperature.
In the improved calculation (for all temperature) the results of which
are depicted in Fig.(2b), the supression of the low frequency
phonon is reduced considerably. \\

\vskip 0.3cm
\noindent {\small Fig 2(b). Temperature dependence of the Raman intensity.} \\
 
\noindent A kink structure appears at a
frequency corresponding to the CDW gap. While this structure is
not effected on lowering the temperature, at high frequencies
the charge carrier scattering decreases with decreasing
temperature. Experimental measurements have shown that on
lowering the temperature the low frequency phonon acquires a
shift and a Fano-line shape, while a new peak emerges in between
the two peaks. The later can be identified with the kink
corresponding to the CDW gap. Furthermore the results of the temperature
expansion calculation has an unphysical divergence of intensity
at zero frequency, which is rectified in the all temperature
calculation as shown in Fig.(2b). The variation of dopant concentration in
these systems can be mimicked by varying the density of states
$N_1(0)$. The variation of the Raman intensity with dopant concentration
$(N_1(0))$ is shown in Fig 3 which shows a large suppression of background
intensity with increasing $N_1(0)$. Similar results for the SC state will 
be published elsewhere \cite{19}. \\

\noindent {\small Fig 3. Variation of the Raman intensity with 
dopant concentration} \\

\noindent The adoptation of the model involving the fluctuating CDW state
and the charge bag superconductivity to the alkali
doped fullerides requires a careful analysis of the Raman data
\cite{6,9}. In this system the Raman spectrum shows several (at
least 10) phonon peaks in the case of the insulating solid
fullerite. Most of these phonons arise from the intramolecular
vibrations of the carbon atoms in $C_{60}$. Two of these phonons
are of $A_g$ symmetry and the rest all have $H_g$ symmetry. On
going over to the alkali doped fulleride $A_{3}C_{60}$ ($A~=~ K,~Rb$) the
system becomes a metal and Raman intensity of most of the $H_g$
modes is suppressed. Which again reappears in the insulating
$A_6C_{60}$ samples. The suppression of the intensity is
attributed to the strong coupling of these phonons to the doped
charge carriers, because of which the phonons acquire a large
width. Since the intramolecular modes are effected, in our
interpretation these should be the zone folded phonons.
Therefore the CDW state is intramolecular in nature
corresponding to the transition from a uniform bond length 
between the carbon atoms to
the stable structure of long and short bonds of the fullerene
molecule. Under this assumption the CDW gap will correspond to
the band gap which is rather large ($\sim$ 1 ev). Since the
Raman continuum appears only above the CDW gap, it will not be
observable in the Raman spectrum, atleast in the frequency
region where the phonons are there. This is clearly depicted in
Fig. 4, which also shows the temperature dependence of the Raman
spectrum in the metallic (doped) state. Note the relative suppression
of the intensities of three of the phonons compared to those in
the insulating state(not shown in figure). In the later state all
the phonons will have the same intensity as there is no self energy
for them.  It is also clear from the figure that on lowering the
temperature (increasing the parameter $c1$) the intensity of
some of the phonons is suppressed considerably. Furthermore, the
suppressed phonons show a hardening of their frequency and
undergo substantial broadening. All these features are in
agreement with the experimental observations. A particular
success of the model is that it provides an explaination as to
why a Raman continuum has not been observed in alkali doped
fullerides. The variation of Raman intensity with
change in the dopant concentration and the coupling constant has also 
been studied and will be reported elsewhere \cite{19}. In the above 
modeling of the fluctuating CDW state for the alkali doped fullerides,
it is unlikely that 
the large CDW gap will ever vanish with doping. \\

\noindent {\small Fig 4. Raman intensity for the alkali doped fullerides.}
\vskip 0.4cm

Hence superconductivity in this system will arise only due to the pairing
of quasiparticles in the upper band as proposed in the original charge
bag model \cite{8}. In this model the calculated Raman intensity \cite{20}
in the SC state does not show much change from that in the normal state.
Again this is in agreement with the experimental observations for the
alkali doped fullerides.
\par  Similar adoptation of the model
to the organic superconductor $(BEDT-TTF)_2I_3$ has been
reported earlier \cite{10}. 
This organic SC
being a 2D-system, the fluctuating CDW state and the charge bag model
has been adopted to it. A zero temperature calculation of the Raman
intensity \cite{10} has qualitatively reproduced the experimental
observations.
\par In conclusion it was shown that most of the features
observed in the Raman data can be understood in terms of the
proposed non-Fermi liquid descriptions of the normal states of
these systems and the corresponding bag mechanism for superconductivity. \\

$\star$ Invited talk presented by one of us (SNB) at the international
conference on SCES-95, Goa, India (1995). \\

\end{document}